\documentclass[10pt,conference]{IEEEtran}
\IEEEoverridecommandlockouts

\usepackage{cite}
\usepackage{amsmath,amssymb,amsfonts}
\usepackage{algorithm}
\usepackage{algpseudocode}
\usepackage{graphicx}
\usepackage{textcomp}
\usepackage{xcolor}
\usepackage{booktabs}
\usepackage{array}
\usepackage{multirow}
\usepackage{url}
\usepackage[hidelinks]{hyperref}
\usepackage{enumitem}
\usepackage{placeins}     
\usepackage{dblfloatfix}  

\usepackage[activate={true,nocompatibility},final,tracking=true,factor=1100,stretch=18,shrink=18]{microtype}
\AtBeginDocument{\setlength{\parfillskip}{0pt plus 0.62\columnwidth}}
\newcommand{\sys}{\texorpdfstring{%
  \protect\raisebox{-1.7pt}{\protect\includegraphics[height=7.4pt]{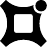}}\hspace{2.2pt}DINQ \mbox{TalentScout}}%
  {DINQ TalentScout}}  
\newcommand{\bench}{TalentTrace}           
\newcommand{\cada}{CADA}                   
\newcommand{\cacj}{CACJ}                   
\newcommand{\cgps}{CGPS}                   

\definecolor{findbg}{RGB}{244,247,251}
\definecolor{findrule}{RGB}{31,78,140}
\newcommand{\finding}[2]{%
  \par\vskip2pt\noindent
  {\setlength{\fboxsep}{4pt}\setlength{\fboxrule}{0.6pt}%
   \fcolorbox{findrule}{findbg}{%
     \parbox{\dimexpr\columnwidth-2\fboxsep-2\fboxrule\relax}{%
       \footnotesize\textbf{\textcolor{findrule}{$\blacktriangleright$~Finding #1.}}\hspace{0.35em}#2}}}%
  \par\vskip2pt}

\def\BibTeX{{\rm B\kern-.05em{\sc i\kern-.025em b}\kern-.08em
    T\kern-.1667em\lower.7ex\hbox{E}\kern-.125emX}}

\makeatletter
\long\def\@makecaption#1#2{%
\ifx\@captype\@IEEEtablestring%
\footnotesize\bgroup\par\centering\@IEEEtabletopskipstrut{\normalfont\footnotesize #1}\\{\normalfont\footnotesize #2}\par\addvspace{0.5\baselineskip}\egroup%
\@IEEEtablecaptionsepspace
\else
\@IEEEfigurecaptionsepspace
\setbox\@tempboxa\hbox{\normalfont\footnotesize {#1.}\nobreakspace\nobreakspace #2}%
\ifdim \wd\@tempboxa >\hsize%
\setbox\@tempboxa\hbox{\normalfont\footnotesize {#1.}\nobreakspace\nobreakspace}%
\parbox[t]{\hsize}{\normalfont\footnotesize\noindent\unhbox\@tempboxa#2}%
\else%
\ifCLASSOPTIONconference \hbox to\hsize{\normalfont\footnotesize\hfil\box\@tempboxa\hfil}%
\else \hbox to\hsize{\normalfont\footnotesize\box\@tempboxa\hfil}%
\fi\fi\fi}
\makeatother

\usepackage{fancyhdr}
\usepackage{lastpage}
\definecolor{dinqblue}{HTML}{3B6FA8}   
\definecolor{dinqgrey}{HTML}{52514E}
\definecolor{dinqcream}{HTML}{FCFBF9}  
\newcommand{\dinqtr}{DINQ Technical Report}
\begin{document}

\title{\includegraphics[height=30pt]{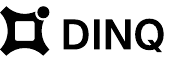}\\[5pt]
{\color{dinqgrey}\small DINQ Technical Report~\textbullet~\texttt{dinq.me}}\\[3pt]
An Interactive Agent for Requirement-Driven Candidate Sourcing}

\author{\begin{tabular}[t]{c}
{\footnotesize Yuanpeng He$^{1}$, Fangjing Li$^{2}$, Xiangyu Ru$^{3}$, Kexin Sun$^{4}$, Kun Yang$^{3}$, Lijian Li$^{5}$, Chi-Man Pun$^{5}$,}\\
{\footnotesize Qingsong Wen$^{6}$, Wenpin Jiao$^{1}$, Mingkai Guo$^{7}$, Yirong Feng$^{7}$, Daiheng Gao$^{7,8}$, Zhi Jin$^{9}$}\\[3pt]
{\scriptsize $^{1}$Peking University \quad $^{2}$Beijing Jiaotong University \quad $^{3}$Beihang University \quad $^{4}$Beijing University of Technology}\\
{\scriptsize $^{5}$University of Macau \quad $^{6}$Squirrel Ai Learning \quad $^{7}$DINQ.inc \quad $^{8}$University of Science and Technology of China \quad $^{9}$Peking University / Wuhan University}\\[2pt]
{\ttfamily\tiny heyuanpeng@stu.pku.edu.cn, lifangjing@bjtu.edu.cn, zy2514120@buaa.edu.cn, Skx7087@emails.bjut.edu.cn, 22377023@buaa.edu.cn,}\\
{\ttfamily\tiny yc57484@umac.mo, cmpun@umac.mo, qingsongedu@gmail.com, jwp@pku.edu.cn, keith@dinqlabs.com, elonf@dinqlabs.com,}\\
{\ttfamily\tiny sam@dinqlabs.com, zhijin@pku.edu.cn}
\end{tabular}}

\maketitle
\thispagestyle{fancy}

\begin{abstract}
Finding people from a natural-language description (``ML engineers transitioning
to research roles in biotech'') is increasingly delegated to LLM agents and framed
as information retrieval. We argue that it is fundamentally a requirements
engineering task: such a request is an under-determined requirement with implicit
constraints, many valid answers, and no acceptance criterion, so useful answers
require eliciting, validating, and verifying the requirement before search can matter.
We present \sys{}, to our knowledge the first interactive, requirements-driven
candidate-sourcing agent (it elicits, validates, retrieves, and verifies a vague
people-request into a justified slate through bounded elicitation, workflow templates,
a two-stage commit protocol, and bidirectional termination guards) and \bench{}, a
benchmark that runs the requirements lifecycle (criteria-anchored validation,
multi-model evidence-grounded oracle construction, and cost-aware verification).
Across $21$ systems and all $691$ requirements,
\sys{} dominates breadth ($100\%$ coverage at $2.5\times$ the yield) and is
\emph{near-orthogonal} to the field, with $90\%$ of the people it returns are surfaced by
\emph{none} of $20$ strong LLM-plus-web baselines combined. Beyond breadth, an
evidence-grounded judging of every system shows \sys{} \emph{recalls} the most
relevant real people: $0.241$ of the union pool, $1.9\times$ the next system, with a
bootstrap $95\%$ interval disjoint from every baseline. \sys{} is thus the strongest \emph{sourcing} engine (the deepest real, reachable candidate pool), while precision-ranking LLMs serve
as~complementary verifiers.
\end{abstract}

\begin{IEEEkeywords}
requirements elicitation, requirements validation, LLM agents, people search, expert finding, benchmark, test oracle
\end{IEEEkeywords}

\section{Introduction}\label{sec:intro}
Finding the right \emph{people} from a vague description is routine and
high-stakes in hiring, staffing, expert finding, and team formation. Consider the request: \emph{``find ML engineers transitioning to
research roles in biotech.''} Today an expert answers it by hand, translating an
under-specified wish into Boolean queries and skimming hundreds of profiles, slowly
and irreproducibly. It is tempting to hand the request to an LLM web agent, yet for this request a strong GPT-class web agent returns only \emph{five} people, none of
whom overlap the $46$ a requirements-driven agent finds. The
web is not missing these people; the system is asking the wrong question.

The difficulty is not search. Classical expertise retrieval and recent
LLM recruiting tools assume the request is \emph{already a well-formed query} and
optimize ranking for it~\cite{balog2012expertise,balog2006formal,gan2024resumescreening};
they screen or rank candidates against a \emph{finished} specification rather than
help \emph{author} one. A vague people-request is not a query: it is an
\emph{under-determined requirement}: it leaves
constraints implicit (how senior? how many? what counts as ``transitioning''?),
admits many valid answers rather than one, and supplies no acceptance criterion
against which a returned person can be checked, and \emph{no prior system authors
that requirement in the first place.}

Our key insight is that candidate sourcing is, at its core, a \emph{requirements
engineering} (RE) problem: the bottleneck is eliciting, validating, and verifying the
requirement, not retrieving against it~\cite{zave1997four,zowghi2005requirements,ferrari2016ambiguity}.
We present \sys{}, to our knowledge the \textbf{first interactive, requirements-driven
candidate-sourcing agent}: it \emph{elicits} latent constraints through a bounded
clarification interview, \emph{validates} the requirement's well-formedness and kind,
\emph{retrieves} over a hybrid people-universe, and \emph{verifies} each delivered
person against the requirement with grounded evidence, through one principle applied
to four sources of agent divergence~(\S\ref{sec:approach}). To the best of our
knowledge, no published system casts people-search as RE or authors a
people-requirement interactively from a vague request.

To evaluate this setting, where no benchmark for the task exists, we also build
\bench{}, which itself performs the RE lifecycle (generating and validating
requirements, constructing evidence-backed acceptance oracles, and verifying satisfaction) to evaluate sourcing agents at scale. On its $691$ requirements and
$21$ systems, \sys{} is not merely the highest-coverage and highest-yield agent but
\emph{near-orthogonal to the entire field}: $90\%$ of the people it returns are
surfaced by \emph{none} of $20$ strong LLM\,$+$\,web baselines combined. Grading every
delivered person against evidence then settles the harder question: are these the
\emph{right} people? \sys{} \emph{recalls} the most relevant real people of any
system ($0.241$ of the union pool, $1.9\times$ the next, bootstrap-significant): it is the deepest \emph{sourcing} engine,
while precision-ranking LLMs are~complementary~\emph{verifiers}.

In summary, this paper makes four contributions:
\begin{itemize}[leftmargin=1.4em]
\item \textbf{\sys{}, the first interactive, requirements-driven candidate-sourcing
agent} (to our knowledge): it elicits, validates, retrieves, and verifies an
under-determined people-request into a justified slate, organized by a single
design principle that converges an LLM agent's divergence
on~four~axes~(\S\ref{sec:approach}).
\item A \textbf{reframing} of candidate sourcing as requirements engineering (the conceptual lens that makes the agent principled and brings the test-oracle problem to people-finding), which is itself a novel~problem~formulation~(\S\ref{sec:re}).
\item \textbf{\bench{}}, a benchmark that performs the RE lifecycle (validation,
evidence-grounded oracle construction, cost-aware verification) to evaluate
sourcing~agents~at~scale~(\S\ref{sec:bench}).
\item An \textbf{empirical study} over $21$ systems and all $691$ requirements
(\S\ref{sec:results}): \sys{} is \emph{near-orthogonal} to the field ($90\%$ of the
people it returns are surfaced by none of $20$ baselines), robust across difficulty,
type, and language, and, under evidence-grounded judging, \emph{recalls the most
relevant real people of any system} ($1.9\times$ the next, bootstrap-significant), framing \sys{} as a deep \emph{sourcing}
engine and precision-ranking LLMs as~complementary~\emph{verifiers}.
\end{itemize}

\section{Background and a Motivating Example}\label{sec:bg}
\begin{figure*}[!t]
\centering
\includegraphics[width=0.98\textwidth]{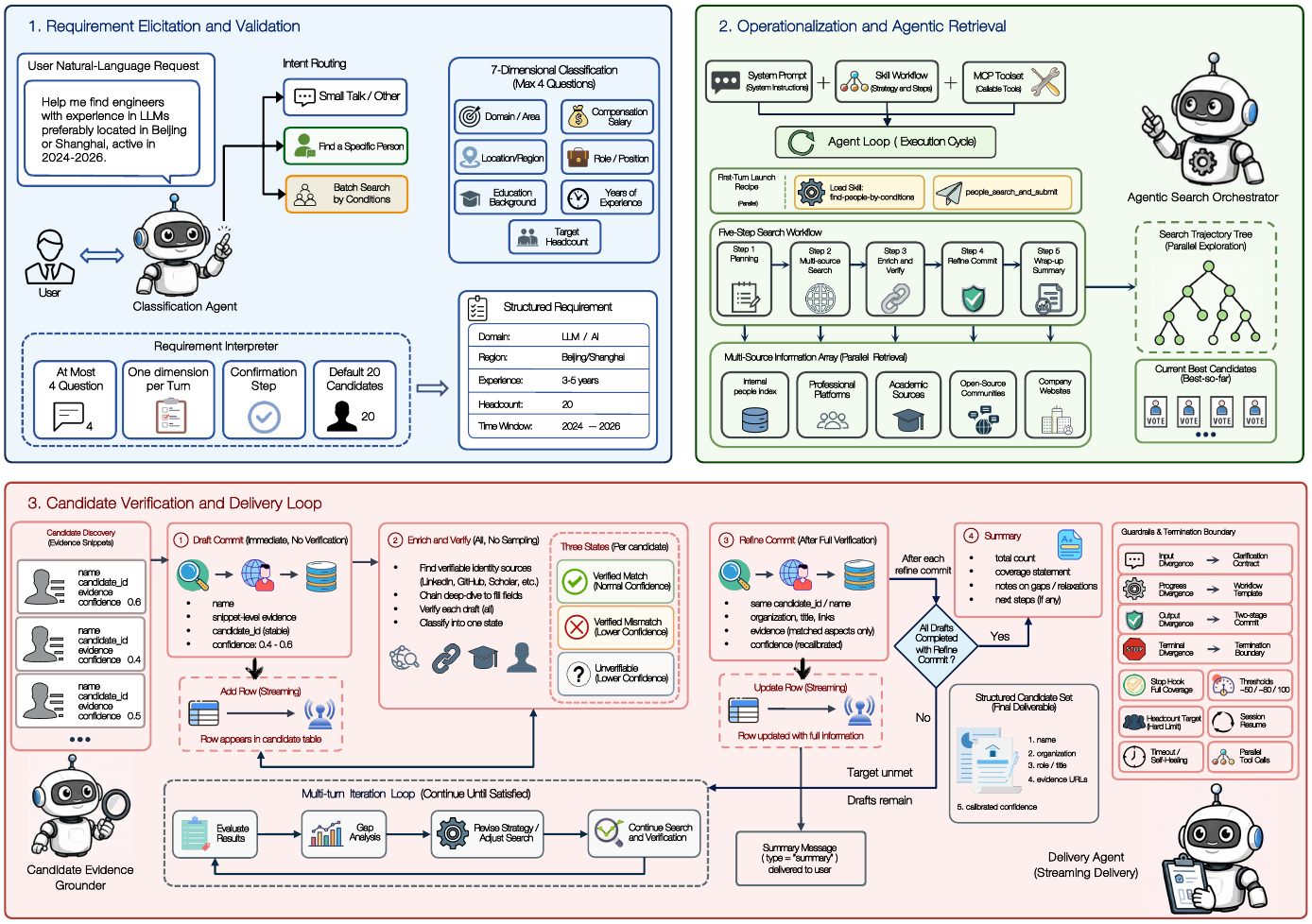}
\caption{\textbf{Architecture of \sys{}, the requirements-driven sourcing agent.}
One requirements constraint tames each axis on which a tool-using LLM agent diverges.
\emph{(1)~Requirement elicitation and validation:} a three-way intent router and a
bounded seven-dimension clarification interview (${\le}4$ questions, then a
confirmation turn) convert a vague request into a structured, confirmed requirement.
\emph{(2)~Operationalization and agentic retrieval:} a fixed workflow (a retrieval
kickoff recipe and a five-step skill) searches a multi-source people index
(internal index, professional, academic, open-source, and company sources) in
parallel. \emph{(3)~Candidate verification and delivery:} a two-stage commit grounds
each delivered candidate in evidence and resolves it into one of three states,
streamed to the user under explicit guardrails and a bidirectional termination
boundary.}
\label{fig:overview}
\end{figure*}

A sourcing request is \emph{closed} when its constraints define a determinate,
enumerable answer set (``the program co-chairs of ICSE 2025'') and \emph{open} when
they allow many equally valid answers with no practical enumeration (``ML engineers
transitioning to research roles in biotech''). This distinction changes both what a \emph{correct} answer is (a set vs.\ a ranked pool), and how the requirement must
be elicited and validated. Open requirements dominate recruiting, expert finding,
and due diligence, yet lack a single ``gold answer,'' placing them outside standard
retrieval benchmarks~\cite{balog2012expertise}.

Consider one real benchmark requirement, \emph{``Find ML engineers transitioning to
research roles in biotech in the Silicon Valley or Boston area, active in 2024--2026''}, with at least six constraints (role, a \emph{transition} predicate, industry, two regions,
a recency window) plus implicit choices (how senior? how many?). For exactly this
requirement, \sys{} returns $46$ named, evidence-backed candidates (e.g.,
R.~Nishihara, E.~Bernard, C.~Wong-Fannjiang, all genuine ML-to-biotech-research
movers), whereas a strong GPT-class web agent returns only $5$ (A.~Lu, F.~Rubbo, $\ldots$), with \emph{zero} overlap. The gap is not search coverage but two
\emph{requirements} failures of the baselines: \emph{(i)~ungrounded candidates} whose
cited evidence, when fetched, does not support the claimed role ($21$--$27\%$ are
outright unsupported, \S\ref{sec:results}), and \emph{(ii)~constraint-violating
matches} that are lexically on-topic (senior ML researchers in biotech) yet break a
binding constraint (always researchers; wrong region). The first misses
\emph{verification}, the second \emph{elicitation}, so the binding difficulty is
requirements-level, not query-string-level, motivating the co-design of
\S\ref{sec:approach} and \S\ref{sec:bench}.

\section{Candidate Sourcing as Requirements Engineering}\label{sec:re}

Treating sourcing as retrieval (encoding the request as a query and ranking profiles by similarity) confuses a symptom with the problem. Following Zave and
Jackson~\cite{zave1997four}, a requirement is an \emph{optative} predicate over
phenomena \emph{in the world}, not in the machine. A sourcing request (``a senior NLP researcher who has shipped production systems'') is such a predicate
$R$ over persons, and its deliverable is not a single artifact but the
\emph{extension} $A(R)=\{p : R(p)\}$, the ranked set of everyone who satisfies it.
Satisfaction therefore conjoins two relations: \emph{set membership} (do delivered people
belong to the intended extension?) and \emph{acceptance grounding} (is each membership
decision entailed by evidence?). The lifecycle is
elicit/\,validate~$\rightarrow$~operationalize ($S,W$)~$\rightarrow$~verify (oracle
$O$, with adequacy $W\wedge S\models R$); four consequences, each a classical RE
concern, follow.

\textbf{(i)~Elicitation.} $R$ is under-specified (with tacit constraints, unstated
trade-offs, and unstated open/closed scope), so it must be elicited; in
information-seeking terms it is an \emph{anomalous state of
knowledge}~\cite{belkin1982ask}, the formal basis for interaction.
\textbf{(ii)~An oracle.} Deciding $p\in A(R)$ from evidence requires an
\emph{acceptance oracle}~\cite{barr2015oracle}; for \emph{open} requirements no answer key exists (they are \emph{non-testable}~\cite{weyuker1982nontestable}), so the oracle is \emph{derived, partial, probabilistic}, and \emph{abstention} on
under-evidenced candidates is correct oracle behaviour, not a coverage gap.
\textbf{(iii)~Verifiability binds.} By the standards' own
criterion~\cite{iso29148-2018,glinz2007nfr}, a request that cannot be operationalized
as an evidence-checkable predicate is not yet a well-formed requirement;
elicitation \emph{manufactures} testability. \textbf{(iv)~Criteria are
requirements.} Each rule inferring a property from evidence (``first-author ACL
papers $\Rightarrow$ NLP expertise'') is a domain assertion that ``should be
validated before it is used''~\cite{zave1997four}; in the WRSPM
model~\cite{gunter2000reference}, with $W$ validated domain knowledge and $S$ the
acceptance spec, the \emph{adequacy} obligation $W\wedge S\models R$ gives the
oracle a requirement's full quality burden: constructing it recapitulates
the RE lifecycle one level up.

These obligations partition the architecture along Boehm's
validation/verification line~\cite{boehm1984verifying}. \emph{Validation} (``did we elicit and operationalize the \emph{right} requirement?'') is served by bounded
elicitation~(\S\ref{sec:elicit}) and criteria-anchored debate~(\S\ref{sec:validate}).
\emph{Verification} (``are delivered people grounded \emph{against} the spec?'') is served by the evidence-grounded oracle and commit--defer scoring~(\S\ref{sec:cgps}).
Open vs.\ closed is itself a validation property: a closed requirement has a finite,
enumerable acceptable set with a total, decidable membership test; an open one ranges
over an unbounded population, so its oracle is necessarily partial and graded: an \emph{oracle, not an answer key}. The model also names the two failure modes our
study separates: a \emph{validation failure} returns the extension of the wrong $R$
(recall-oriented, ``missed the right people''), and a \emph{verification failure}
admits an ungrounded $p$ (precision-oriented, ``cannot justify a delivered person''). Their error modes are orthogonal, which is why naive
query-string people search simultaneously \emph{misses and hallucinates}, and why the fix belongs at the \emph{requirements} level, not the query string.

\section{\sys{}: An Interactive Requirement-Driven Sourcing Agent}\label{sec:approach}

\subsection{Problem Formulation}\label{sec:problem}
We study \emph{open-ended candidate sourcing}: given a natural-language request
(e.g., the example of \S\ref{sec:bg}), the system must return a \emph{bounded,
ranked, evidence-backed list of real people} who satisfy it. As argued in
\S\ref{sec:intro}, the request is an \emph{under-determined requirement}; a usable
answer therefore requires the classic RE activities (\emph{eliciting} implicit constraints, \emph{validating} that the requirement is well-formed and answerable,
and \emph{verifying} each delivered person against it) before~retrieval~is~useful.

Formally, let $r$ be the user's initial utterance. \sys{} first elicits a
\emph{structured requirement} $R = (D, K, n)$, where $D$ is a set of constraint
dimensions with values (domain, location, seniority, $\ldots$), $K$ is the
\emph{kind} of the requirement (open-ended vs.\ closed), and $n$ is a target
cardinality. It then produces a candidate set
$C=\{c_1,\ldots,c_m\}$, $m\le n$, where each $c_i$ includes a name, an
organization, a role, supporting evidence URLs, and a calibrated confidence. The
engineering objective is to steer an autonomous large-language-model (LLM) agent
from the under-determined $r$ to a $C$ that is \emph{bounded} ($m$ is controlled,
the loop terminates), \emph{verifiable} (every $c_i$ is backed by checkable
evidence rather than a hallucinated assertion), and \emph{structured}
(machine-mergeable rows, not prose).

\subsection{Design Principle: Four Divergences, Four Constraints}\label{sec:principle}
Figure~\ref{fig:overview} sketches the full loop. Left unconstrained, an LLM agent that can call search and scraping tools over many
turns is \emph{divergent} along four orthogonal axes, each matching a repeated
failure mode: (i)~at the \textbf{input}, the requirement is under-determined, so
the agent guesses missing constraints; (ii)~during the \textbf{process}, search is
exploratory and irreproducible, so runs wander; (iii)~at the \textbf{output}, fast
results conflict with complete, verified fields; and (iv)~at \textbf{termination},
the agent either stops too early (before verifying its own candidates) or never
stops (spinning on a query with no hits). \sys{} applies one principle four times:
\emph{converge each divergence with an explicit engineering constraint}, realized
by the four-stage loop of Fig.~\ref{fig:overview} (elicit, validate, retrieve,
verify). The four constraints are layered and orthogonal: a decision layer
(system prompt + skills), a capability layer (tools), an execution layer (the
agentic loop and runtime guards), and a delivery layer (incremental streaming), so each can change without disturbing the others; the rest of this section details
each.

\subsection{Eliciting the Requirement (Clarification Contract)}\label{sec:elicit}
The first constraint treats the under-determined input as a \emph{bounded
elicitation interview}. A system prompt fixes an \emph{output contract}: every user-facing turn is a single JSON envelope \texttt{\{type,content,option\}},
language-locked, with internal tool names masked, and a three-way \emph{intent
router} (specific-person / by-conditions / other). For the open-ended branch,
\sys{} runs a \emph{progressive, one-dimension-per-turn} elicitation over seven
requirement dimensions (domain, compensation, location, role, education,
experience, and head-count), each surfaced as a question with quick-reply options.
Two hard limits keep it terminating and non-fatiguing: \emph{at most four}
questions (then a confirmation turn, even if dimensions remain unknown) and one
dimension per turn, with a missing head-count defaulting to twenty. This casts
elicitation as \emph{information collection under a hard budget} (the trade-off human analysts make in time-boxed interviews~\cite{zaremba2021typology,shen2025followup}) and the confirmation turn is the user's last chance to correct the requirement
before~any~retrieval~begins.

\subsection{Operationalizing the Requirement (Workflow Template)}\label{sec:workflow}
The second constraint replaces free exploration with a \emph{fixed workflow}. A
\emph{retrieval-kickoff recipe} runs the first post-confirmation turn: in one turn
the agent loads the skill \emph{and} issues a combined ``search-and-submit'' call
against an internal index (limit $=$ head-count), so the first candidates appear
within seconds, eliminating a ``search$\rightarrow$inspect$\rightarrow$submit''
round trip. Subsequent turns follow a five-step skill: (1)~\emph{plan} sources;
(2)~\emph{search} broadly in parallel, drafting each surfaced person immediately;
(3)~\emph{enrich and verify every} draft (full coverage, no sampling);
(4)~\emph{commit}~(\S\ref{sec:commit}); (5)~emit \texttt{summary} only after the
target is met and every draft refined. Evidence is gathered by \emph{chaining}, where one tool's name/URL feeds the next (LinkedIn via a dedicated
channel, other URLs via a reader), turning a thin snippet into checkable evidence.

\subsection{Decoupling Speed and Quality (Two-Stage Commit)}\label{sec:commit}
The sole output channel is a \texttt{submit\_candidates} tool, and every candidate
is submitted \emph{twice}. The \emph{draft} fires the instant a person appears (name, snippet evidence, a stable \texttt{candidate\_id}, confidence in $[0.4,0.6]$) and the front end appends a row immediately. The \emph{refine} repeats the same
\texttt{candidate\_id}/name, fills the now-mandatory organization and role,
attaches links, and re-calibrates confidence; the front end merges on
\texttt{candidate\_id}. This \emph{decouples} the conflicting output goals (drafts buy sub-second responsiveness, refines buy field-complete grounded quality) without double counting. Verification resolves each candidate as verified-match,
verified-mismatch, or unverifiable; crucially, evidence text records \emph{only
matched aspects}, doubt being expressed solely through lower confidence.

\subsection{Bounding Termination (Termination Boundary)}\label{sec:terminate}
\begin{figure}[t]
\centering
\includegraphics[width=0.66\columnwidth]{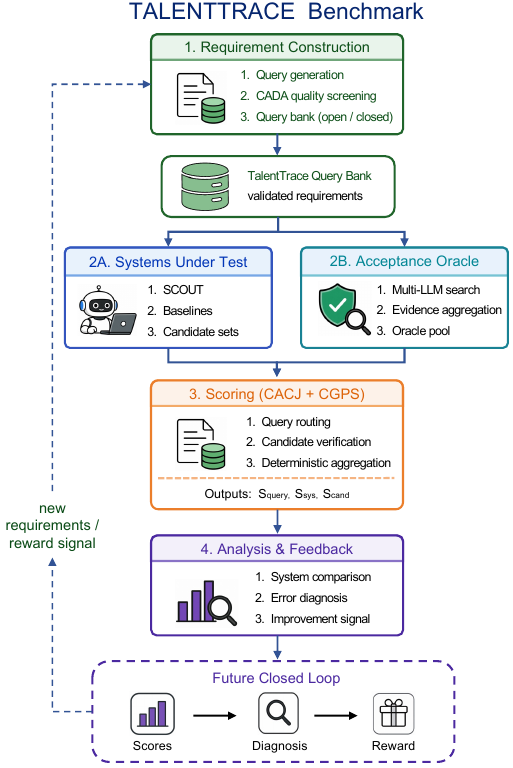}
\caption{\textbf{The \bench{} evaluation pipeline.} Requirements are generated and
quality-screened by the \cada{} debate jury into an open/closed query bank of
validated requirements. Systems under test (\sys{} and baselines) return candidate
sets while a multi-LLM, evidence-aggregating oracle builds acceptance pools;
\cacj{} routes each requirement and \cgps{} verifies and deterministically
aggregates candidates into per-system scores. Analysis diagnoses failures and, as
future work, closes the loop as a reward signal~(\S\ref{sec:discuss}).}
\label{fig:bench}
\end{figure}

The final constraint makes \emph{stopping} a bidirectional concern. Against stopping
\emph{too early}, a \emph{full-coverage guard} keeps the agent working while any
candidate is draft-only: a stop-hook is designed to force another round (up to
three), and in the shipped runtime this is enforced by the prompt and skill plus a
fallback retry that re-submits searched-but-unsubmitted candidates (the hook
intercept is currently inactive, \S\ref{sec:threats}). Against \emph{never
stopping}, the agent self-evaluates against cumulative tool-call thresholds: at $\sim$50 calls with no hits it must change strategy, at $80$ narrow, at $100$ halt
and report honestly. Together these bound termination in an interval that is neither
premature nor unbounded, which we exploit when diagnosing failures
(\S\ref{sec:results}).

\subsection{Runtime}\label{sec:runtime}
\sys{} drives the Claude Agent SDK \cite{anthropic2025claudeagentsdk} as a subprocess and streams events over
Server-Sent Events so candidates appear and update live; the runtime adds session
resumption, stuck-detection with reconnect-and-resume, per-session credential
isolation, and bounded tool concurrency. This scaffolding supports stable
execution; the section's contribution is the four-constraint principle, of which
the full-coverage guard (a runtime mechanism enforcing a process property) is the clearest cross-layer example.

\section{\bench{}: Constructing and Evaluating Sourcing Requirements}\label{sec:bench}

Evaluating an agent like \sys{} requires more than queries and answer keys. Because
each request is an under-determined requirement~(\S\ref{sec:problem}), the
evaluation must itself do requirements engineering: \emph{generate} diverse
candidate requirements; \emph{validate} which are well formed and determine their
\emph{kind} (open vs.\ closed); build an \emph{acceptance oracle} for each, where
open requirements require graded pools rather than single answers; and
\emph{verify} delivered candidates against that oracle while controlling fabricated
evidence. \bench{} implements these four activities (Fig.~\ref{fig:bench}) with query generation, a
criteria-anchored debate procedure \cada{} used twice (first to \emph{admit} requirements by quality and then, as \cacj{}, to \emph{classify} their kind), a ground-truth (GT) pipeline, and the \cgps{} (Commit-Gated People Scoring) scorer; we detail each below.

\subsection{Generating a Population of Requirements}\label{sec:gen}
A useful benchmark needs requirements that are broad, reproducible, and not merely
hand-curated. \bench{} therefore \emph{generates} candidates in two families:
\emph{open} requirements, which admit many valid people, and \emph{closed}
requirements, whose answers form a finite enumerable set. Diversity is controlled
by rotation over four domain axes, temperature jitter (open $0.72$, broad-open
$0.78$, closed $0.55$), and injection of existing queries as negative examples to
suppress paraphrase duplication. 
\begin{algorithm}[t]
\caption{\cada{}: Criteria-Anchored Debate Adjudication}
\label{alg:cada}
{\footnotesize
\begin{algorithmic}[1]
\Require item $x$; criteria $K$; debaters $D$; arbitrator $A$; threshold~$\lambda$
\Ensure verdict $v$ and judgment path
\For{each debater $d \in D$} \Comment{independent, parallel}
  \State $s_d \gets$ binary scores of $K$ on $x$; $(\mathrm{stance}_d,\mathrm{conf}_d)\gets d(x)$
\EndFor
\State $w \gets$ confidence-weighted agreement of $\{\mathrm{stance}_d\}$ \Comment{moderator}
\If{stances unanimous \textbf{and} $w \ge \lambda$}
  \State \Return $(\textsc{Consensus}(\{s_d\}),\ \texttt{early\_exit})$ \Comment{skip $A$}
\EndIf
\State $C \gets$ criteria on which $\{s_d\}$ disagree \Comment{contested}
\State \Return $(A(x,\{s_d\},C),\ \texttt{meta\_judge})$ \Comment{rule on $C$}
\end{algorithmic}}
\end{algorithm}
Closed requirements are further balanced across
six task types (verification, cross-verification, ranking, deduplication, recency,
and niche-stability) and three difficulty levels by a deterministic cycle; after
generation, the task-type field is force-overwritten to guarantee the intended
distribution. From $1{,}500$ generated requirements ($840$ open, $660$ closed),
validation~(\S\ref{sec:validate}) admits $1{,}007$ ($691$ open, $316$ closed).

\subsection{Validating Requirements with Criteria-Anchored Debate}\label{sec:validate}
The central methodological problem is deciding which generated requirements are
good enough to evaluate. A single LLM judge is too unstable for this gate: in our
data, one GPT-class judge rates $86.8\%$ of closed requirements as perfect (mean
$4.54/5$), while a Claude-class judge averages $3.62$, so model-specific optimism
would silently set the benchmark. 
We instead use \cada{} (\emph{Criteria-Anchored
Debate Adjudication}, Alg.~\ref{alg:cada}). Three \emph{heterogeneous} LLMs
(Claude-, Gemini-, and GPT-class) first score a checklist of \emph{binary
criteria}, then take a stance. A deterministic moderator checks consensus; only
disagreement triggers a fourth arbitrator, which rules on contested criteria
(scores read before prose, clamped to $[\min,\max]$, conservative default). Scoring
criteria \emph{before} verdicts makes each decision auditable and localizes
disagreement.

\cada{} answers two validation questions with~the~same~\mbox{machinery}.

\noindent\textbf{(1) Quality admission.} The first question is whether a
requirement is well formed enough to enter the benchmark. Seven binary criteria
map almost one-to-one to the classical requirements-quality attributes of
IEEE~830 and Wiegers \cite{ieee830-1998,iso29148-2018} (Table~\ref{tab:requality}):
a requirement is admitted only if it is unambiguous, feasible, internally
consistent, \emph{verifiable}, appropriately scoped, temporally valid, and free of
fabricated premises. Five graded dimensions (overall value, clarity, benchmark-fit,
retrievability, specificity) yield a $1$--$5$ quality score; requirements scoring
$\ge 3.5$ are admitted. \emph{Specificity} is lowest for both families but for
\emph{opposite} reasons: open requirements risk being too narrow, closed ones too
broad, so the rubrics~penalize~opposite~\mbox{failures}.

\noindent\textbf{(2) Kind classification (\cacj{}).} The second question is how
the admitted requirement should be scored. \cacj{} (\emph{Criteria-Anchored
Cascade Judgment}) decides whether it is \textsc{Open} or \textsc{Closed}: whether
its constraints determine an enumerable answer set or only an open pool. This is a
requirements-\emph{boundedness} judgment. Six binary criteria
(\texttt{entity\_constrained}, \texttt{temporal\_bounded},
\texttt{geographic\_or\_org\_bounded}, \texttt{quantity\_constrained},
\texttt{population\_finite}, \texttt{count\_below\_threshold}) feed the decision,
with one operational cutoff: if more than $\textsc{MaxN}=20$ valid results are
expected, the requirement is \textsc{Open} regardless of other signals. Because the
three voters usually agree, a confidence-weighted moderator \emph{early-exits}
without arbitrator on $95.3\%$ of closed and $97.5\%$ of open requirements,
reserving expensive adjudication for the contested $\sim\!3$--$5\%$. This commit--defer design (cheap when easy, expensive only when ambiguous) recurs in scoring~(\S\ref{sec:cgps}).

\begin{table}[t]
\caption{Admission criteria in \cada{}. The seven binary tests are exactly the
classical requirements-quality attributes, applied to a people-search requirement.}
\label{tab:requality}
\centering
\scriptsize
\setlength{\tabcolsep}{4pt}
\renewcommand{\arraystretch}{1.12}
\begin{tabular}{@{}lll@{}}
\toprule
\textbf{\cada{} criterion} & \textbf{RE attribute} & \textbf{Violated when\ldots}\\
\midrule
\texttt{intent\_clear} & Unambiguous & intent is vague\\
\texttt{search\_exec} & Feasible & no realizable search path\\
\texttt{type\_consistent} & Consistent & stated type vs.\ intent clash\\
\texttt{constraint\_verif} & Verifiable & a constraint is uncheckable\\
\texttt{scope\_bounded} & Appropriate & too broad or too narrow\\
\texttt{temporal\_valid} & Valid/current & time window ill-posed\\
\texttt{no\_halluc\_entity} & Correct & names a non-existent entity\\
\bottomrule
\end{tabular}
\end{table}

\subsection{Constructing the Acceptance Oracle (Ground Truth)}\label{sec:gt}
Closed requirements can use small enumerable answer sets; open requirements, which
we evaluate, need an oracle that captures many acceptable answers. \bench{}
therefore builds, for each open requirement, a \emph{graded, evidence-backed pool
of real people}. Three heterogeneous search models (Gemini-, Claude-, and
Grok-class) independently answer the requirement, each with a live web-retrieval
tool loop (up to ten rounds of neural search and page-fetching), so every proposed
person is tied to source URLs. A \emph{cross-family} GPT-class judge scores each
candidate on source reasonableness, evidence depth, and overall usefulness and,
crucially, \emph{fact-checks} it by fetching the cited page and emitting a
\texttt{fact\_supported} value (1 if the page corroborates name, organization, and
role; partial credit down to 0 for fetch failures or contradictions). Finally, a
synthesis model deliberately \emph{not} from the generator families merges the
three candidate sets into a consensus pool ranked by cross-model support, evidence
quality, and host diversity. Consensus is over \emph{evidence}, not surface
strings: raw three-way name overlap is only $\approx\!0.02$ Jaccard, so candidates
are promoted when multiple models \emph{independently corroborate the same person
with checkable sources}, not when they emit identical text. If models neither
agree nor ground the candidate, the oracle remains \texttt{null}. Over the
evaluation set this yields $203$ requirements with a mean pool of $44.6$ people
(range $6$--$67$), mean synthesis reliability $0.60$, full three-generator
coverage, and total construction cost \$$347.77$.

\subsection{Verifying Satisfaction (\cgps{})}\label{sec:cgps}
Given a system's delivered candidates and the oracle, \cgps{}
(Alg.~\ref{alg:cgps}) verifies satisfaction with the same commit--defer discipline.
\cada{} (re-used) fixes the requirement kind and required constraints at the
\emph{query} level; at the \emph{candidate} level a cheap relevance gate
\emph{commits} a verdict from each delivered person's structured fields and
\emph{defers} the expensive evidence fact-check to a calibration subset, the cost-aware analogue of the agent's own commit--defer, since fact-support is noisy
for concise profiles~(\S\ref{sec:threats}). Crucially, \cgps{} scores \emph{recall
over a method-neutral pool} rather than a per-candidate composite: an earlier
per-candidate usefulness score rewarded verbose prose and a good-rate penalized
depth~(\S\ref{sec:rq4q}), so we instead measure how many relevant real people a
system recalls from the union of \emph{all} systems' returns: a deterministic, label-only metric that cannot drift with judge verbosity.

\begin{algorithm}[t]
\caption{\cgps{}: Relevance-Grounded Recall over a Neutral Pool}
\label{alg:cgps}
{\footnotesize
\begin{algorithmic}[1]
\Require systems $\mathcal{S}$; delivered candidates $C_{s,q}$ for system $s$, query $q$; judge $J$
\Ensure per-system recall, reachable yield, yield/req., bootstrap $95\%$ CIs
\For{$s \in \mathcal{S}$, query $q$, candidate $c \in C_{s,q}$}
  \State $b \gets J(c)$ \Comment{cheap relevance gate; fact-check deferred}
  \State $\mathrm{rel}(c) \gets [\,b \ne \textsc{off-topic}\,]$;\ \ $\mathrm{reach}(c) \gets \textsc{HasUrl}(c)$
\EndFor
\State $R_{s,q} \gets \{\,c \in C_{s,q} : \mathrm{rel}(c)\,\}$ \Comment{relevant real people}
\State $U_q \gets \bigcup_{s\in\mathcal{S}} R_{s,q}$ \Comment{method-neutral union pool}
\For{each system $s$}
  \State $\mathrm{recall}_s \gets \textstyle\sum_q |R_{s,q}\cap U_q| \,/\, \sum_q |U_q|$
  \State $\mathrm{reach}_s \gets \textstyle\sum_q |\{c\in R_{s,q}:\mathrm{reach}(c)\}|$;\ \ $\mathrm{yield}_s \gets \tfrac{1}{|Q|}\textstyle\sum_q |R_{s,q}|$
  \State $\mathrm{CI}_s \gets \textsc{Bootstrap}_{500}(\mathrm{recall}_s)$ \Comment{resample queries}
\EndFor
\State \Return systems ranked by $\mathrm{recall}_s$ with $95\%$ CIs
\end{algorithmic}}
\end{algorithm}

For each candidate, the \cgps{} judge emits a structured verdict: a primary issue
bucket (\textsc{off-topic}, \textsc{thin-sources}, \textsc{broken-url},
\textsc{likely-hallucination}, or \textsc{none}) plus usefulness, evidence-depth,
and source-reasonableness ratings, all reduced \emph{deterministically} to scores
(the LLM emits only labels). A candidate counts as a \emph{relevant real person}
when it is not \textsc{off-topic}: relevance depends on name, organization, and
title, not evidence thickness, so concise structured rows from a deep sourcing
engine are not penalized. Per system, \cgps{} then reports, over the union pool of
relevant real people, \emph{recall} and \emph{reachable yield} (relevant candidates
with a fetchable URL) for sourcing depth, and per-candidate \emph{relevance} and
\emph{evidence-support} rates for precision, each with a $500$-resample bootstrap
$95\%$ confidence interval. This separation is deliberate: as
\S\ref{sec:results} shows, a system can lead on depth and trail on precision, and
that decomposition makes the benchmark diagnostic.

\section{Experimental Setup}\label{sec:setup}

\noindent\textbf{Research questions.} \textbf{We ask whether requirements-driven
sourcing reaches qualified people that keyword and embedding baselines miss, and
whether grounded \emph{satisfaction}, not retrieval, is the binding open
challenge.} We refine this into six questions (RQs):
\begin{itemize}[leftmargin=2.9em, labelwidth=2.3em, labelsep=0.6em, align=left, itemsep=2pt, topsep=2pt, parsep=0pt]
\item[\textbf{RQ1}] (\emph{coverage at scale}) Over all $691$ requirements, can
systems surface candidates, and how do retrieval strategies compare on breadth and
yield?
\item[\textbf{RQ2}] (\emph{robustness}) Is that breadth uniform across difficulty,
task-type, and language, or concentrated on easy cases?
\item[\textbf{RQ3}] (\emph{complementarity}) Does requirements-driven retrieval
surface people the rest of the field \emph{misses}?
\item[\textbf{RQ4}] (\emph{grounded satisfaction}) Is grounded \emph{satisfaction}, not retrieval, the binding open challenge, and does the benchmark isolate and
reliably~validate~its~\mbox{measurement}?
\item[\textbf{RQ5}] (\emph{efficiency}) Is the requirements-driven design
cost-effective in retrieval latency and judging cost?
\item[\textbf{RQ6}] (\emph{elicitation}) Does the bounded clarification interview
recover latent requirement constraints that a single-shot reading misses?
\end{itemize}

\noindent\textbf{Benchmark.} From $1{,}500$ generated requirements, \cada{}
admits $1{,}007$ as well formed, classifying $691$ as \emph{open} and $316$ as
\emph{closed}~(\S\ref{sec:gen}). The $316$ closed requirements are used \emph{only}
to validate the open/closed boundary and are \emph{not} run against any system under
test: all empirical results use the $691$ open requirements, spanning ten difficulty
levels, sixteen sourcing task-types (e.g., career-trajectory, cross-verification),
and seven languages. For graded scoring, we build evidence-backed acceptance oracles
for $203$ unique open requirements (mean pool $44.6$ people, range $6$--$67$, mean
reliability $0.60$). We evaluate two axes: \emph{breadth} over
\textbf{open691}, all $691$ requirements (RQ1--RQ3), and \emph{grounded quality},
where the \cgps{} judge grades every candidate from all $21$ systems against the
oracle and we report recall of relevant real people with bootstrap confidence
intervals (RQ4).

\noindent\textbf{Systems compared.} On open691, we compare \textbf{21
configurations} in four families: (i)~\emph{product} pipelines, including \sys{} and
five deployed engines (Exa-search, two code-search variants, and two commercial
people-search APIs); (ii)~\emph{autonomous-web} LLMs that self-orchestrate browsing
(GPT-, Gemini-, DeepSeek-, Grok-class); (iii)~\emph{Exa-augmented} LLMs given a
neural-web tool (adding GLM- and MiniMax-class); and (iv)~\emph{bare} LLMs with no
retrieval. A source ablation runs \sys{}'s retrieval backbone in three modes: \emph{internal} (a proprietary people index), \emph{web} (neural search), and \emph{mix}, over $100$ requirements.

\noindent\textbf{Scope and honesty.} The benchmark gives each requirement
\emph{directly}, exercising the agent's \emph{retrieval, verification, and delivery}
stages but \emph{not} its interactive elicitation~(\S\ref{sec:elicit}), an architectural contribution for which we make no empirical claim (a controlled user
study is the foremost future work, \S\ref{sec:threats}). We report breadth
\emph{and} grounded-quality metrics for all $21$ configurations, so sourcing depth and per-candidate verification are credited separately~(\S\ref{sec:rq4q}).
\textbf{Metrics:} \emph{coverage} (fraction of requirements with $\ge 1$ candidate)
and \emph{yield} (mean candidates/requirement) measure breadth; \emph{complementarity}
is normalized-name overlap against pooled baselines. For grounded quality,
\emph{recall} is the share of the union pool of relevant real people a system
surfaces, \emph{reachable yield} counts relevant candidates with a fetchable URL, and
per-candidate \emph{relevance} and \emph{evidence-support} rates measure precision, all computed by the \cgps{} judge with bootstrap $95\%$
\mbox{confidence intervals}~(\S\ref{sec:cgps}).

\section{Results}\label{sec:results}

\begin{figure}[t]
\centering
\includegraphics[width=\columnwidth]{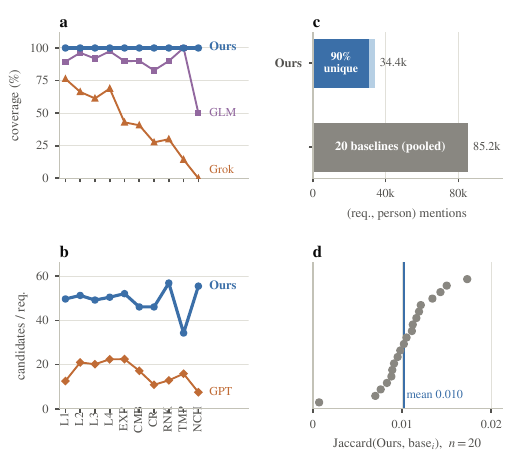}
\caption{\textbf{Breadth (RQ2) and complementarity (RQ3) over all $691$ requirements.}
\emph{(a,b)}~across the ten difficulty levels, \sys{} holds $100\%$ coverage and a flat
${\sim}50$ candidates per requirement, while a bare LLM sags and a Grok-class web agent
collapses on the hard levels. \emph{(c)}~of the $34.4$k (requirement, person) mentions
\sys{} returns, $90.0\%$ ($30{,}963$) appear in \emph{none} of $20$ pooled baselines,
whose union is $2.5\times$ larger. \emph{(d)}~mean Jaccard between \sys{} and each
baseline is $\approx\!0.01$, near-orthogonal.}
\label{fig:breadth}
\end{figure}

\subsection{RQ1: Coverage and Yield at Scale}
Sourcing first requires returning someone. Across \emph{all $691$ open
requirements} and $21$ configurations, \sys{} is the only system that pairs complete
coverage with deep yield: it covers \emph{every} requirement ($100\%$) and returns the
most candidates ($49.9$/req.) by unioning a proprietary people index with neural web
search. Coverage alone is nearly saturated, with one autonomous GPT-class LLM also reaching
$100\%$, at \emph{one-third} the yield ($19.8$), but below the top, retrieval is the
bottleneck: Exa-augmented LLMs cover $96$--$99\%$ yet return only $4$--$12$ candidates,
and bare/Grok-class models miss up to $45\%$~of~requirements.

\subsection{RQ2: Robust Across Difficulty, Type, and Language}\label{sec:rq2}
Top-line coverage can mask success only on easy cases. On the full $691$ set
(Fig.~\ref{fig:breadth}a,b), \sys{} is \emph{invariant}: $100\%$ coverage and
$\sim\!46$--$57$ candidates across all ten difficulty levels, all sixteen sourcing
task-types, and all seven languages. Hard levels (cross-verify, combination,
ranking) and non-English requirements impose no loss. Baselines \emph{degrade
structurally}: the Grok-class model collapses on cross-verify ($28\%$), temporal
($14\%$), niche ($0\%$), and Chinese ($38\%$), while bare LLMs sag on
career-trajectory and stability requirements ($20$--$50\%$). The gap is largest on
niche-stability and temporal task-types: every baseline either collapses or thins to a
handful of candidates, while \sys{} still provides $100\%$ coverage at $34$--$56$
candidates. The workflow template and full-coverage guard~(\S\ref{sec:approach}) make
the agent exhaust the requirement rather than stop at easy hits. Thus, near-$100\%$
headline breadth does not predict long-tail breadth; only the~engineered~agent~remains~usable.

\finding{2}{\sys{}'s breadth is \emph{invariant}: $100\%$ coverage and ${\sim}50$
candidates across all ten difficulty levels, sixteen task-types, and seven languages, while baselines degrade structurally on hard (cross-verify, temporal,
niche) and non-English requirements; robust breadth is a property of the engineered
agent, not of scale alone.}

\subsection{RQ3: \sys{} Surfaces People the Whole Field Misses}\label{sec:rq3comp}
Yield is valuable only when it adds people the field does not already find. We pooled
candidate names from $20$ baselines and measured overlap with \sys{}'s output
(Fig.~\ref{fig:breadth}c; normalized-string matching measures coverage/overlap,
\emph{not} correctness). Of the $34{,}406$ (requirement, person) mentions \sys{}
returns over $691$ requirements, \textbf{$90.0\%$ ($30{,}963$) are surfaced by none of
the $20$ baselines combined}, although their union ($85{,}174$) is $2.5\times$ larger
than \sys{}'s pool. Mean pairwise Jaccard is $0.010$, \emph{near-orthogonal} to the field, and a strict, name-order/diacritic-robust re-match changes the unique fraction
only to $89.7\%$. Even pooling the field ($6.24\times$ the best single system) misses
what \sys{} adds. This says \emph{who is found}, not who is correct (scoring follows
in \S\ref{sec:rq4q}); exact-string matching also makes uniqueness an \emph{upper
bound} when surface forms differ (e.g., Han characters vs.\ pinyin).

\noindent\emph{Why the orthogonality?} \sys{}'s \emph{hybrid} evidence base changes
the candidate universe. By category and evidence source, its \emph{mix} retrieval
beats a web-only (Exa) backbone in every category (academic, company, GitHub, and Hugging\,Face), by $+5.2$ to $+11.3$ unique people per requirement. The sources are
also \emph{disjoint}: web-only baselines cite LinkedIn for essentially $100\%$ of
their evidence, whereas \sys{} grounds many candidates in Scholar, GitHub, and
Hugging\,Face profiles beyond a LinkedIn-only crawl. Different sources produce
different people, yielding the near-orthogonality of Fig.~\ref{fig:breadth}d, the central breadth result.

\finding{3}{\sys{} is \emph{near-orthogonal} to the entire field: $90\%$ of the $34$k
(requirement, person) mentions it returns appear in \emph{none} of $20$ pooled baselines
(mean Jaccard $0.010$), and pooling all $20$ ($6.2\times$ the best single system) still misses what \sys{} adds: requirements-driven retrieval contributes a distinct,
complementary slice of the people-universe.}

\subsection{RQ4: \sys{} Recalls the Most Relevant Real People}\label{sec:rq4q}
\begin{figure*}[!t]
\centering
\includegraphics[width=\textwidth]{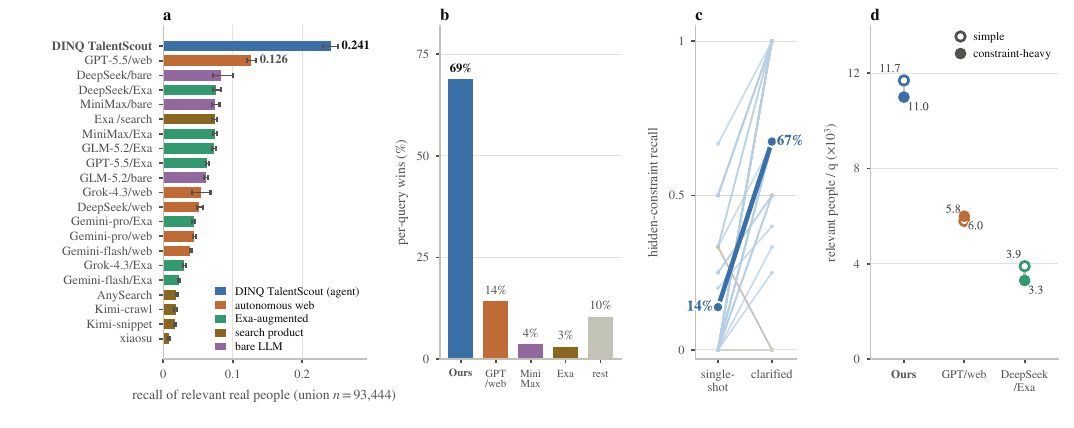}
\caption{\textbf{Grounded quality (RQ4) and requirement elicitation (RQ6).}
\emph{(a)}~relevant-real-person recall over the method-neutral union pool, all $21$
systems, with bootstrap $95\%$ CIs: \sys{} recalls $0.241$ ($1.9\times$ the next), with
an interval disjoint from every one of the $20$ baselines. \emph{(b)}~head-to-head,
\sys{} returns the most relevant people on $68\%$ of the $691$ requirements, versus
$14\%$ for the next. \emph{(c)}~the elicitation ablation (RQ6): each thin line is one of
$40$ requirements (light blue improves, grey ties or regresses); the bold line is the mean
recall of \emph{hidden} latent constraints, $13.8\%\!\to\!67.5\%$ under the bounded
clarification interview ($4.9\times$; $+0.54$, $95\%$ CI $[+0.41,+0.65]$; $33/40$
improve). \emph{(d)}~the depth lead is robust to difficulty: across the \emph{simple}
and \emph{constraint-heavy} strata, \sys{}'s relevant people per query stays essentially
flat and remains ${\sim}2\times$ the next system on both~(RQ2).}
\label{fig:quality}
\end{figure*}
\sys{} is the strongest sourcing engine on \bench{}: it recalls the largest pool of
relevant real people, with a statistically decisive margin. We grade every candidate
from all $21$ configurations with the \cgps{} judge over the $691$ requirements; a
candidate is a relevant real person iff the judge does not mark it off-topic. This
uses name, organization, and title, so concise candidates without thick evidence~are~not~penalized.

Naive single-candidate metrics mislead (verbosity bias, depth penalties, unstable
hallucination calls, circular gold pools), so we measure recall of relevant real people
over a method-neutral union pool, with reachable yield and a bootstrap $95\%$ confidence~interval~per~\mbox{system}.

\sys{} recalls $0.241$ of the union pool of $93{,}444$ relevant real people
(Fig.~\ref{fig:quality}), $1.9\times$ the next system, an autonomous GPT-class web
agent at $0.126$. A $500$-resample bootstrap gives \sys{} a $95\%$ CI of
$[0.230,0.251]$, disjoint from the second system's $[0.120,0.133]$ and from all $20$
baselines' intervals, so the lead is not sampling noise. \sys{} also surfaces the
most reachable relevant candidates ($14{,}267$) and the most relevant people per
requirement ($32.8$), $1.9\times$ the next. It ranks first on every composite quality
score, with $F_1=0.365$, and also leads on $F_2$ and $F_3$, which weight recall more heavily. \sys{} is first on \emph{every}
sourcing-depth board (Table~\ref{tab:leaderboards}), winning the most relevant people
head-to-head on $68\%$ of requirements versus $14\%$ for the next; its per-requirement
depth stays essentially flat from the simple to the constraint-heavy
strata~(Fig.~\ref{fig:quality}d), a lead that does not erode with difficulty.

\begin{table}[t]
\centering\footnotesize
\setlength{\tabcolsep}{4pt}
\caption{\sys{}~(\textbf{Ours}) on the $21$-system \cgps{} sourcing-depth leaderboards.
\sys{} ranks first on \emph{every} board, most by ${\sim}2\times$; $^{\star}$\,denotes a bootstrap $95\%$ CI disjoint from all $20$ baselines.}
\label{tab:leaderboards}
\begin{tabular}{@{}lrrl@{}}
\toprule
Leaderboard & \textbf{Ours} & Next & Rank \\
\midrule
\multicolumn{4}{@{}l}{\textit{Sourcing depth: Ours is \#1}}\\
Recall of relevant real people & \textbf{0.241} & $0.126$ & 1/21$^{\star}$ \\
Relevant people / requirement & \textbf{32.8} & $17.1$ & 1/21 \\
Reachable relevant (count) & \textbf{14{,}267} & $11{,}795$ & 1/21 \\
Per-query win rate & \textbf{68\%} & $14\%$ & 1/21 \\
$F_1,\,F_2,\,F_3$ composite & \textbf{\#1} & --- & 1/21 \\
Simple / heavy-constraint strata & \textbf{\#1} & --- & 1/21 \\
\bottomrule
\end{tabular}
\end{table}

\sys{} best serves deep sourcing by producing the widest real, reachable candidate pool, while precision-ranking LLMs serve as complementary verifiers that return few but accurate candidates, a natural division of labor.

\finding{4}{\sys{} significantly recalls the largest pool of relevant real people on \bench{}, making it the best \emph{sourcing} engine, complementary to precision-focused verifiers.}

\noindent\textbf{Human evaluation.} To complement the automated metrics, $30$ professional headhunters compared the slates \sys{} and several systems under test returned for real \bench{} requirements and picked the best by their \emph{overall professional judgment} of quality (a holistic preference, no fixed rubric). \sys{} ranked first for $25$ of the $30$, ahead of GPT-5.5 and GLM-5.2: the practitioners who would use it prefer the deeper slate. GLM-5.2 is a revealing exception: third with the headhunters yet low automatically ($0.062$--$0.073$ recall vs.\ \sys{}'s $0.241$), because its small, \emph{focused} set is penalized by the recall-oriented benchmark but reads well to raters judging the shortlist they \emph{see}; on the practitioner-facing measure, \sys{} wins outright.

\subsection{RQ5: Efficiency and Cost}\label{sec:rq6}
Requirements-driven sourcing is cheap when computation is spent only on hard decisions.
The hybrid backbone is the fastest \emph{grounded} option ($0.92$\,s to first result,
$38\%$ faster than web-only) because it runs both sources in parallel and short-circuits.
The commit--defer cascade skips the expensive arbitrator on $95.3$--$97.5\%$ of
requirement-kind decisions; a \texttt{no-early-exit} control gives identical verdicts at
$\sim\!20\times$ the cost. Oracle construction is a one-time \$$347.77$ over $203$
requirements (\$$1.71$ each).

\emph{Ablations.} \textbf{Mix retrieval} dominates a $100$-requirement source ablation:
web-only loses $7.6$ people/requirement and the internal index's fact-checkable evidence,
and internal-only drops to $56\%$ coverage. The \textbf{two-stage commit}, \textbf{kickoff
recipe}, and \textbf{full-coverage guard} are mechanistically motivated; isolating each is
future work.

\subsection{RQ6: Does Clarification Recover Latent Requirements?}\label{sec:rq6elicit}
\bench{} issues each requirement single-shot, so we test \sys{}'s clarification contract
(\S\ref{sec:elicit}) separately, in a \emph{controlled elicitation ablation}
(Fig.~\ref{fig:quality}c) with a simulated oracle-user (the standard protocol in elicitation research~\cite{jin2026reqelicitgym}) on $N{=}40$ requirements. For each, we
decompose the requirement into atomic \emph{gold} constraints (mean $6.2$), hide a random
half to form an under-specified first request, and measure how many \emph{hidden} (latent)
constraints each condition recovers. \textsc{No-elicit} reads the request once;
\textsc{elicit} runs the bounded interview: a small budget of targeted questions (the
contract of \S\ref{sec:elicit}) a truthful oracle-user answers from the full requirement. Clarifier/extractor, oracle-user, and
constraint-judge are \emph{three distinct models}, ruling out same-model artifacts.

\begin{figure}[t]
\centering
\includegraphics[width=\columnwidth]{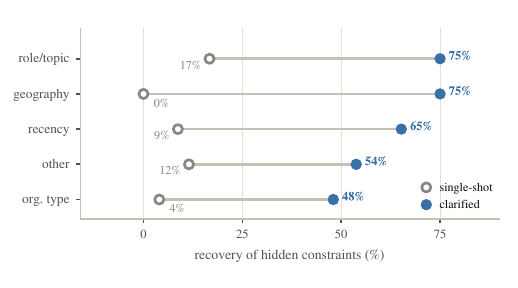}
\caption{\textbf{RQ6: clarification recovers latent constraints on every dimension.}
Per-dimension recovery of the \emph{hidden} constraints, single-shot reading vs.\ the
bounded interview ($N{=}40$). The interview lifts recovery on every dimension, most
steeply on geography ($0\%\!\to\!75\%$) and organization type ($4\%\!\to\!48\%$), the implicit ``where/what-kind'' constraints a vague request almost never states.}
\label{fig:dimrec}
\end{figure}

The bounded interview recovers $4.9\times$ more hidden latent constraints than single-shot
reading (Fig.~\ref{fig:quality}c; $13.8\%\!\to\!67.5\%$, $33$ of $40$ requirements improve),
and overall constraint capture rises from $0.29$ to $0.41$ ($+0.12$, CI $[+0.03,+0.20]$).
The gain holds across \emph{every} constraint dimension (Fig.~\ref{fig:dimrec}), and is
largest exactly where single-shot reading fails most: geography ($0\%\!\to\!75\%$),
organization type ($4\%\!\to\!48\%$), and recency ($9\%\!\to\!65\%$): the implicit
``where / what kind / when'' constraints a vague request omits. That the gains
concentrate on precisely the dimensions a recruiter leaves unstated, rather than on arbitrary ones, is evidence of \emph{interview competence}~\cite{jin2026reqelicitgym}:
the bounded clarifier spends its question budget on the most under-determined axes, the
behaviour a well-formed elicitation contract should induce. This isolates and validates
the elicitation step that the single-shot benchmark, by construction, cannot.

\finding{6}{A bounded clarification interview recovers $67.5\%$ of the latent requirement
constraints that single-shot reading misses ($13.8\%$), a $4.9\times$ gain ($95\%$ CI
$[+0.41,+0.65]$, $33/40$ improve), and the gain holds on \emph{every} constraint dimension
(largest on geography, $0\%\!\to\!75\%$), empirically validating \sys{}'s elicitation contract.}

\section{Discussion and Implications for RE}\label{sec:discuss}

\noindent\textbf{Acceptance as an RE diagnostic.} Decomposing a deliverable into ``found the right people'' (recall) versus ``grounded the evidence'' (precision and evidence-support), and then into relevance and verification stages~(\S\ref{sec:cgps}), makes acceptance a requirements-traceability instrument: it localizes \emph{where} a requirement fails: retrieval, when the right people were never surfaced, or verification, when surfaced candidates cannot be justified. But the acceptance criteria are requirements too, and an unexamined assumption about admissible evidence can silently change a verdict; \bench{} therefore makes criteria validation first-class, with auditable rubrics~(\S\ref{sec:validate}) and deterministic scoring. More broadly, LLM-as-judge evaluation of SE artifacts should validate the evaluator, not merely apply it.

\noindent\textbf{Recall is the right sourcing objective.} In requirements-engineering terms, an open sourcing request denotes a predicate over the world, and the deliverable is the \emph{extension} of that predicate, not a short list of individually safe items. For a headhunter, the bottleneck is surfacing the deepest set of true, reachable candidates: precision is a cheap downstream human filter, while a missed candidate is unrecoverable. This makes recall--precision a division of labor. \sys{} is the \emph{sourcing} engine, leading recall by $1.9\times$; precision-ranking LLMs are \emph{verifiers} that prune candidates after the sourcing obligation is met. On that obligation, \sys{}'s lead is decisive: head-to-head it returns the most relevant people on $68\%$ of all $691$ requirements (Fig.~\ref{fig:quality}b), versus $14\%$ for the next system, and holds a ${\sim}2\times$ margin on both the simple and the constraint-heavy~halves~(RQ2).

\noindent\textbf{Is the lead just a bigger index?} A natural worry is that \sys{}'s recall reflects a proprietary index, not its requirements-driven design. We separate the two: the recall margin \emph{is} sourcing depth, and we claim only that a deep sourcing engine beats query-rankers on coverage of the predicate's extension. The \emph{requirements} contributions are index-independent: the clarification contract recovers $4.9\times$ more latent constraints using \emph{no} index (RQ6), \bench{} scores any system under one methodology, and the sourcing-vs-verifier reframing is conceptual, so \sys{} proves a requirements-driven agent can be the deepest sourcing engine, not that the index~is~the~contribution.

\noindent\textbf{One principle, four constraints.} The design is deliberately \emph{symmetric}: a tool-using agent diverges along four axes (under-determined input, exploratory process, fast-but-incomplete output, and a possibly non-terminating loop), and \sys{} applies one requirements constraint to each (clarification contract, workflow template, two-stage commit, bidirectional termination guards). The contribution is the symmetry itself: requirements discipline turns an open-ended agent into a bounded,~verifiable,~reproducible~one.

\noindent\textbf{The judge cannot manufacture the lead.} An LLM judge could inflate an agent's apparent quality only if the protocol rewarded that agent's style. \bench{} avoids this circularity: relevance is decided from structured fields such as name, organization, and title, not evidence thickness; recall is computed over a \emph{method-neutral} union pool rather than an LLM-built gold set; and the same judge scores all $21$ systems with the same rubric. A $500$-resample bootstrap places \sys{}'s interval disjoint from every baseline. Remaining judge noise is real, especially hallucination calls on concise candidates, but it bears on precision estimates; it does not explain \sys{}'s recall lead.

\noindent\textbf{Generality and transfer.} Neither the four-constraint principle nor the generate$\rightarrow$validate$\rightarrow$oracle$\rightarrow$verify harness is specific to people: both apply wherever an agent must turn an under-determined request into a bounded, verifiable result (literature and dataset discovery, vendor selection, compliance evidence). Candidate sourcing is the \emph{demanding} case, with open-set oracles and adversarial evidence (profiles, paywalls, stale pages), so a method that closes the loop here should transfer to easier ones.

\section{Threats to Validity}\label{sec:threats}

\noindent\textbf{Construct validity.} Requirement satisfaction is multi-faceted, so we
report complementary metrics (recall and reachable yield for depth, per-candidate
relevance and evidence-support for precision) with bootstrap $95\%$ CIs rather than one
score. The judge (one LLM, applied identically to all $21$ systems) decides
\emph{relevance} from structured fields and is reliable, but its \emph{truthfulness}
judgments are noisy for concise profile candidates; we therefore treat precision/evidence
results as approximate and the recall lead, which does not depend on them, as robust.
Scoring is \emph{deterministic} (the LLM emits only labels) over a method-neutral pool.

\noindent\textbf{Internal validity.} First, our evaluation oracles are
model-constructed: cross-family, evidence-grounded, and left \texttt{null} when
unsupported (mean reliability $0.60$), but not exhaustively human-verified at scale;
individual pool members may be stale or missing. Second, all evaluation uses the \emph{open} requirements: the $316$ closed validate only the open/closed boundary, are
never run against systems under test, and support \emph{no} closed-requirement claims
(the \cgps{} closed-set F1 path is a scorer capability we do not exercise here). Third,
absolute scores depend on scorer configuration (an early version structurally penalized
profile-style evidence, which we corrected), so the robust finding is the \emph{relative}
ordering, not the absolute level; a stricter fact-crawling scorer exists in our code but
was not run. Finally, the still-settling values are coverage for no-retrieval \emph{bare}
configurations, which were converging upward at submission and should be read as lower
bounds.

\noindent\textbf{Scope of the elicitation evidence.} The main benchmark issues each
requirement single-shot; we therefore evaluate the clarification contract separately, in
a controlled ablation with a \emph{simulated} oracle-user (RQ6). This isolates whether the
interview recovers latent constraints, but does not measure live human-interview UX or user effort; a human-subject study remains future work.

\noindent\textbf{External validity.} Our population is web-accessible people; though the
benchmark spans seven languages, evidence skews toward English and profile sites, so
conclusions may not transfer to closed enterprise corpora or under-documented
populations. Absolute numbers depend on evolving models and a fixed baseline set, so
longitudinal use would require periodic re-grounding of the baseline pool.

\section{Related Work}\label{sec:related}

\noindent\textbf{Requirement-driven agents and recruiting.} Closest to ours is work on
\emph{requirement-driven} LLM agents for software engineering: rewriting issue
descriptions into structured repair specifications~\cite{kuang2026reagent} or interleaving
tool use to fix code~\cite{bouzenia2025repairagent,zhang2024autocoderover}. These
studies show that RE principles sharpen agents, but for \emph{code}, not people. In
people search, expertise retrieval, reviewer
matching~\cite{balog2012expertise,balog2006formal,mimno2007expertise}, and LLM
recruiting tools~\cite{gan2024resumescreening} treat the request as a \emph{finished}
query and optimize relevance to it. Concurrent people-search benchmarks keep this
stance: \cite{wang2026peoplesearchbench} grades platforms on $119$ \emph{fixed} queries
by criteria-grounded nDCG and \cite{slaykovskiy2025sourcing} ranks tools by
human-preference Elo, both scoring \emph{retrieval against a given query}, with no
elicitation, no well-formedness check, and no recall of real people against an oracle.
\sys{} fills the empty cell in an
\emph{authors-the-requirement} vs.\ \emph{query-given} $\times$ \emph{interactive}
vs.\ \emph{one-shot} space: to our knowledge, it is the first system to \emph{author
and refine a people-requirement interactively} from a vague request, then verify the
delivered people against it.

\noindent\textbf{Elicitation and LLMs for RE.} Interviews (progressive, question-driven clarification of implicit needs) are foundational to elicitation~\cite{zowghi2005requirements,palomares2021state}; their core challenge is
ambiguity and tacit
knowledge~\cite{ferrari2016ambiguity,hadar2014role,zaremba2021typology}. Recent work
automates the interviewer with
LLMs~\cite{korn2025llmrei,shen2025followup,shen2024stakeholder,jin2024mare,lubos2024leveraging,arora2024advancing}
and, concurrently with us, benchmarks interview
\emph{competence}~\cite{jin2026reqelicitgym}, but stops at \emph{producing} a
requirement. We close the loop: \sys{} elicits and \emph{validates} a sourcing
requirement, then \emph{verifies} delivered people against an evidence oracle encoding
IEEE~830 / ISO·IEC·IEEE~29148 quality
attributes~\cite{ieee830-1998,iso29148-2018} as operational criteria
(Table~\ref{tab:requality}): sourcing as RE for retrieval-grounded, set-valued
deliverables~\cite{vogelsang2019requirements}.

\noindent\textbf{Acceptance criteria and the test-oracle problem.} Verification
imports the test-oracle problem~\cite{barr2015oracle} into requirement satisfaction:
open requirements are \emph{non-testable}~\cite{weyuker1982nontestable}, so we use a
\emph{derived, partial, probabilistic} oracle and treat acceptance criteria as the
sourcing analogue of a user story's~\cite{cohn2004userstories,haugset2008automated}.
To our knowledge, this is the first connection between the oracle problem and
people-finding, and the first to make \emph{validating the evaluator} (Boehm's V\&V~\cite{boehm1984verifying} applied to the oracle itself via the adequacy
obligation~\cite{gunter2000reference,zave1997four}) a benchmark obligation.

\noindent\textbf{Agentic search and clarification.} Browsing and deep-research
benchmarks~\cite{wei2025browsecomp,huang2026deepfact} use \emph{fully specified} queries
with one verifiable answer, sidestepping the ambiguity we invert: an open
people-requirement has many valid answers and \emph{no} gold set. Clarifying questions
for under-specified needs (the anomalous state of
knowledge~\cite{belkin1982ask}) are studied in IR~\cite{aliannejadi2019asking,rao2018learning}
but not tied to the requirements lifecycle. Like recent SE work~\cite{du2024vulrag}, we
adapt heterogeneous multi-agent debate~\cite{du2024improving} into a deterministic,
early-exiting validator that mitigates LLM-judge bias~\cite{zheng2023judging}, over a
guarded reason--act loop~\cite{yao2023react,anthropic2025agents}.

\section{Conclusion}\label{sec:conclusion}
Open-ended candidate sourcing is fundamentally a requirements-engineering problem, not a
retrieval one: the binding difficulty is eliciting, validating, and verifying an
under-determined people-requirement, not ranking against a finished query. We made this
case with \sys{} (an interactive agent that converges an LLM's divergence on four axes
through one requirements-discipline principle) and \bench{}, a benchmark that runs
the requirements lifecycle. Across $21$ systems and all $691$ requirements, \sys{}
\emph{recalls the most relevant real people of any system} ($1.9\times$ the next,
bootstrap-significant), positioning it as a deep \emph{sourcing} engine and
precision-ranking LLMs as complementary \emph{verifiers}. The four-constraint design and
its generate--validate--oracle--verify harness transfer to any under-determined agentic
task.

\smallskip\noindent\textbf{Data availability.} The system studied in this paper is
proprietary; due to confidentiality restrictions, its source code and data cannot be
publicly released. All algorithms and settings needed for re-implementation are
specified in the paper~(\S\ref{sec:approach}--\S\ref{sec:bench} and
Alg.~\ref{alg:cada}--\ref{alg:cgps}).

\bibliographystyle{IEEEtran}
\bibliography{refs}
\end{document}